\documentclass[journal]{IEEEtran}
\ifCLASSINFOpdf
\else
\fi
\usepackage{graphicx}
\usepackage{amssymb}
\usepackage{cite}
\usepackage{color}
\usepackage{amsmath,amsfonts,algorithmic}
\begin{document}
%
\title{Identification of head impact locations, speeds, and force based on head kinematics}
%
%
%

\author{Xianghao Zhan, Yuzhe Liu, Nicholas J. Cecchi, Jessica Towns, Ashlyn A. Callan, Olivier Gevaert, Michael M. Zeineh, David B. Camarillo
\thanks{Xianghao Zhan, Nicholas Cecchi, Jessica Towns, Ashlyn Callan, David Camarillo are with Department of Bioengineering, Stanford University, Stanford, CA, 94305, USA; Yuzhe Liu is with School of Biological Science and Medical Engineering, BeiHang University, Beijing, 10019, China; Olivier Gevaert is with Department of Biomedical Data Science, Stanford University, Stanford, 94305, CA, USA; Michael Zeineh is with Department of Radiology, Stanford University, Stanford, 94305, CA, USA.
(Corresponding Author: Yuzhe Liu e-mail: yuzheliu@buaa.edu.cn)}
\thanks{This research was supported by the National Natural Science Foundation of China Young Scholar Program 1230021530, Young Elite Scientists Sponsorship Program by CSTAM 2023-XSC-HW3,  Fundamental Research Funds for the Central Universities YWF-23-Q-1029, the Pac-12 Conference’s Student-Athlete Health and Well-Being Initiative, the National Institutes of Health (R24NS098518), Taube Stanford Children’s Concussion Initiative and Stanford Department of Bioengineering.}}

%
%

\markboth{}%
{Shell \MakeLowercase{\textit{et al.}}: Identification of head impact locations, speeds, and force based on head kinematics}
%



\maketitle

\begin{abstract}
Objective: Head impact information including impact directions, speeds and force are important to study traumatic brain injury, design and evaluate protective gears. This study presents a deep learning model developed to accurately predict head impact information, including location,  speed, orientation, and force, based on head kinematics during helmeted impacts. 
Methods: Leveraging a dataset of 16,000 simulated helmeted head impacts using the Riddell helmet finite element model, we implemented a Long Short-Term Memory (LSTM) network to process the head kinematics: tri-axial linear accelerations and angular velocities. Results: The models accurately predict the impact parameters describing impact location, direction, speed, and the impact force profile with $R^2$ exceeding 70\% for all tasks. Further validation was conducted using an on-field dataset recorded by instrumented mouthguards and videos, consisting of 79 head impacts in which the impact location can be clearly identified. The deep learning model significantly outperformed existing methods, achieving a 79.7\% accuracy in identifying impact locations, compared to lower accuracies with traditional methods (the highest accuracy of existing methods is 49.4\%). 
Conclusion: The precision underscores the model’s potential in enhancing helmet design and safety in sports by providing more accurate impact data. Future studies should test the models across various helmets and sports on large in vivo datasets to validate the accuracy of the models, employing techniques like transfer learning to broaden its effectiveness.
\end{abstract}


\IEEEpeerreviewmaketitle

\section{Introduction}
Head impacts may lead to Traumatic Brain Injury (TBI), which is the neurological disease with the highest global incidence and prevalence \cite{1}. Recent studies found evidence showing that neurological pathologies were closely associated with the brain deformation produced by head impacts \cite{2,3,4,5,6}. Therefore, measuring head impact is essential in TBI research. Mechanically, a head impact is characterized by the impact location on the head, the direction, speed of the impact mass, and impact force. These head impact parameters are influential to the risk of TBI. Obviously, higher impact speed leads to more severe brain deformation and thus results in higher TBI risk. Then, the impact direction and location are also important because the complicated anisotropic brain structures make the brain's vulnerability vary significantly when the head acceleration direction is changed \cite{7}. 

Helmets are designed according to the head impact information: if a helmet is used for a situation where the head impact is of high impact energy, the energy absorbing system is stiffer and heavier, and vice versa. Then, if the risky head impact in the situation frequently happens at specific locations, those locations should be strengthened. For example, position-specific features were found in concussive head impacts in American Football, therefore position-specific helmets were suggested to reduce the TBI risk \cite{8}. The energy absorption at the forehead of the motorbike helmet was strengthened by inserting aluminum honeycomb fillers \cite{9}. Furthermore, specific impact locations were required in the helmet tests \cite{10}, and the scores evaluating the helmet protection were weighted according to the frequency of concussive head impacts. Head impact information is also used to analyze the head impact and identify the TBI risk factors. In American football, front, and side head impacts are more frequent, and the back impact is always more severe \cite{11}, especially for quarterbacks. The side and rear head impacts were found to be associated with obvious performance decrement in on-field American Football games \cite{12}.

Although the head impact information is important, it is hard to directly measure because the sensors have to cover all the potential impact locations. A recent attempt is the Head Impact Remote Sensing (HIRS) system \cite{13}, which is a self-powered sensor array based on a multiangle triboelectric nanogenerator (MA-TENG). HIRS was integrated into the snow helmet to detect head impact and has exhibited a promising ability to measure the impact force and location. However, HIRS applicability in on-field games needs to be further examined considering the weight and expense, and there is no head impact dataset collected by HIRS so far. Another solution is to reconstruct the head impacts with dummies in labs according to the videos: 54 professional football head impacts were reconstructed by Sanchez et al. \cite{14}  and the National Football League (NFL) helmet testing protocols were developed based on the head impact information collected by this method. Furthermore, the helmets that were damaged in head impacts were collected to reconstruct the head impact accident by comparing the cracks in the helmets \cite{15,16,17}. These two methods do not need pre-deploy sensors, but it is limited by two factors: the accuracy of reconstructing head impact according to helmet cracks and the expense of collecting large amounts of damaged helmets. 

An alternative method is to estimate the head impact information according to the head kinematics, which has been widely measured by wearable inertial sensors such as instrumented mouthguards, helmet-based and skin-based sensors \cite{18,19,20,21}.  Based on the head kinematics, mechanical models were developed to reconstruct the head impact and thus retrieve the impact location.  In the Head Impact Telemetry System (HITS), the direction of linear acceleration at the Center of Gravity (CoG) of the head was considered as the impact direction, and part of the head in the opposite direction was considered as the impact region \cite{22}. Then, the direction of angular motions (acceleration, velocity, and location) was used to find an orthogonal cross vector, which is used to correct the impact direction \cite{23}. However, this correction was found not to be effective because only 37\% of the impact was identified as the right of the impact region \cite{23}, and only 28\% of the impact was identified correctly by directly using linear acceleration \cite{22}. Furthermore, another model is developed by neglecting the neck and assuming that the head is a free rigid body \cite{24}. They used linear and angular accelerations to calculate the total force and torque on this body, which was then used to solve the impact direction, location, and force magnitude. Then, a multi-degree of freedom (MDOF) system of head-neck systems was proposed to estimate the impact force, and promising accuracy was found compared with the impact test using Hybrid III headform and neck \cite{25}. However, the accuracies of these kinematics-based methods \cite{24,25} were not analyzed. It should be noted that these kinematics-based methods listed above \cite{22,23,24,25} did not include helmets, but some \cite{22,23,24} were widely used in helmeted sports like American Football. As helmets are designed to disperse the impact force and mitigate the head acceleration, they may generate considerable error in estimating the head impact information. Another limitation of these kinematics-based methods is that they only provide the impact locations, while they can not predict the impact speed and impact force.

In this paper, we developed a new impact information retrieval model based on deep learning to predict all the head impact information including impact location, direction, speed, and impact force according to the head kinematics in helmeted impacts. A large dataset of helmeted head impacts (n=16,000) was established to develop the model, and promising accuracy of head impact information was found. Furthermore, we used an on-field American football dataset of head kinematics with recorded video (n=118) to examine the accuracy of the deep-learning model in on-field games. The impact retrieval model significantly improves the accuracy in identifying the impact location compared with current methods \cite{22,23,24}. With this paper, we published the large head impact simulation datasets (n=16,000) for those willing to develop signal-based data-driven models.

\section{Materials and Methods}

\subsection{Dataset of in-silico Head Impacts}
\begin{figure}

    \centering
    \includegraphics[width=\linewidth]{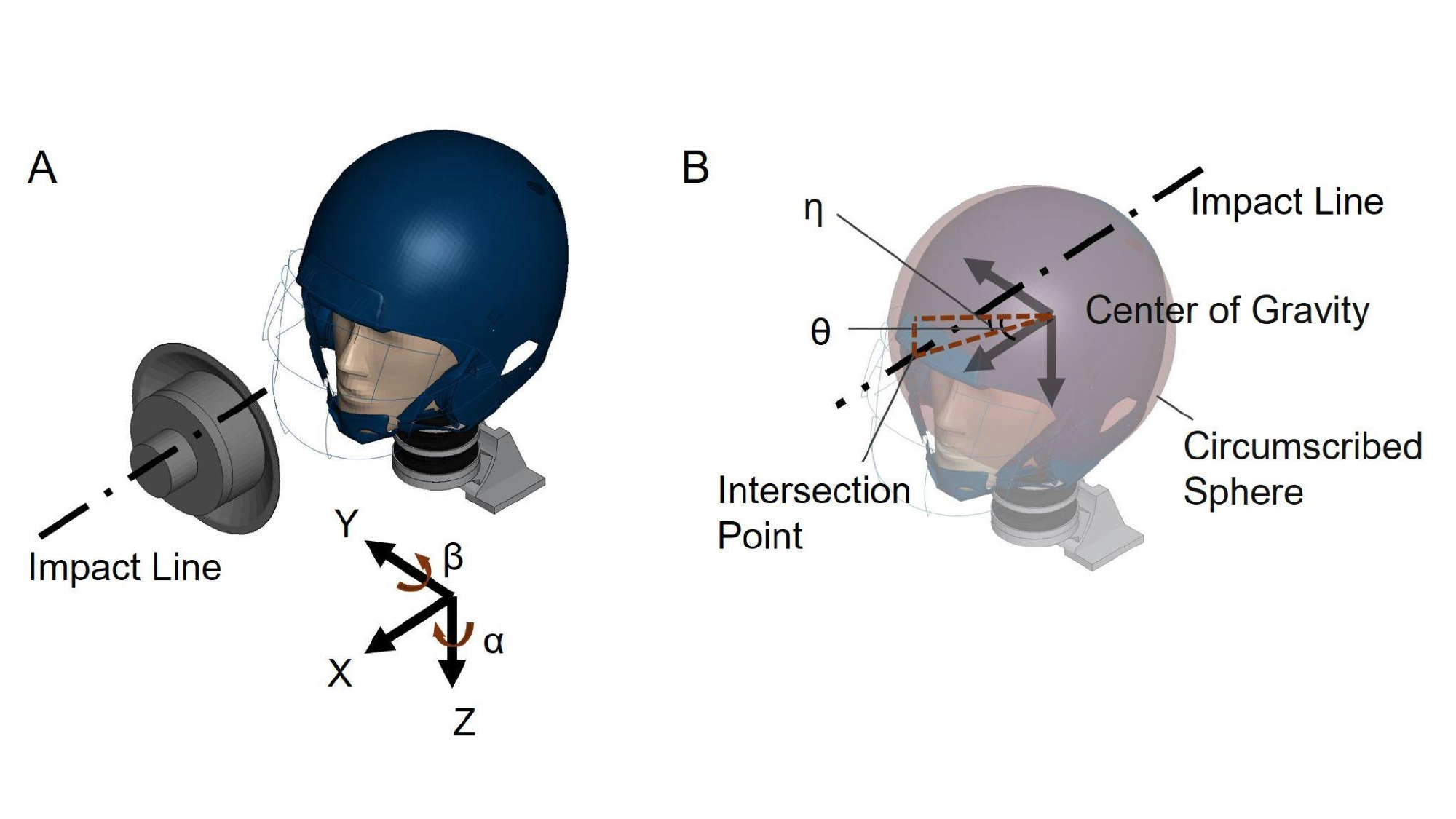}
    \caption{The impact model setup. (A) Finite element model of helmet impact test \cite{26}. The parameters of headform adjustment and the positive directions. The impact line indicates the trajectory of the impactor head; (B) The intersection point between the impact line and the circumscribed sphere.}
    \label{fig1}
\end{figure}

A large head impact dataset with 16,000 head impacts was generated to develop the impact retrieval model based on the numerical model of the NFL standard helmet testing platform consisting of a pneumatics impact, hybrid III headform and neck, and a sliding table to support the neck (Fig. \ref{fig1}). The FE model of Riddell helmets \cite{26} is installed on the FE Hybrid III headform. As the linear impactor fires, the impactor head will be launched at a controlled initial speed along its axial direction until it blows the headform. In both the physical model and the FE model, the farthest distance that the impactor can go is fixed, and we removed this constraint to make sure the impactor can go far enough, which is closer to the actual situation of on-field head impacts. The accelerometer and gyroscope built in the FE headform model will record the linear acceleration and the angular velocity of the headform, and the impact force on the helmet and the head will be recorded. 

The axis of the pneumatics impactor is fixed, and the headform can move translationally and rotationally to have different impact directions and locations. As shown in Fig. \ref{fig1}A, the headform can be adjusted in five degrees of freedom to produce different head impacts: pitch angle, which shows the headform rotation in the sagittal plane, is denoted as $\beta$; yaw angle, which shows the headform rotation in the horizontal plane, is denoted as $\alpha$; the headform translational movement along the impactor axis, denoted as X; the headform translational movement in the horizontal plane and vertically to the impactor axis, denoted as Y; the headform translational movement upwards and downwards, denoted as Z. The roll angle is not a degree of freedom because the whole test setup is axis symmetrical to the impactor axis. Then, since the impactor head can move freely in X direction, a group of parameters including $\alpha$, $\beta$, Y, Z and impact speed determine a head impact: $\alpha$, $\beta$ determine the impact direction, and $\alpha$, $\beta$, Y, Z together determine the impact line, which is the line the impact header moves along. The intersection point between the impact line and the helmet is the impact location. To simplify the calculation of the intersection point, the helmet profile was represented by a sphere with a radius of 135 mm and the center at the head CoG. In the spherical coordinate, the impact location can be represented by $\theta$ and $\eta$, as shown in Fig. \ref{fig1}B.

To avoid the impactor hitting the bottom of the headform, which is impossible in the real world, we excluded $\beta$ which is close to 0° or 180°.  The basic dataset was established with $\alpha$ ranging from 10° to 180° (8 values, $\alpha$ is positive here because of the symmetry), $\beta$ ranging from -45° to 70° (8 values), Y ranging from -120 mm to 120 mm (5 values) and Z ranging from -120 mm to 120 mm (5 values). Impact speed ranges from 3 m/s to 10 m/s with 5 values. These ranges include most of the head impacts in the real world and 8,000 head impacts were generated.

Every head impact was simulated for 150 ms to make sure the impact was finished. Then, linear acceleration, angular velocity, and force on the helmet and head/face were outputted with an interval of 1 ms and then filtered by zero-phase digital filtering with a cut-off frequency of 300 Hz. The headform kinematics were transferred from the anatomical reference (moving with the headform) to the global reference using quaternary. Because the whole impact setup is symmetrical in the y direction, another 8,000 head impacts can be obtained by mirroring head impacts against the vertical plane passing the impactor axis. For the mirrored cases, -1 was multiplied to y and $\alpha$ for an impact setup, the linear acceleration and force in Y direction (X, Y, Z is the direction in the headform reference, X is posterior to anterior, Y is left to right, Z is top to bottom), and angular velocity in X and Z direction. As a result, we generated 16,000 head impacts dataset. 

\subsection{Dataset of on-field American Football Head Impacts and Impact Location Review}
A dataset of head kinematics was collected in the varsity Stanford American Football team with instrumented mouthguards \cite{28,29}. The instrumented mouthguards were built with a set tri-axial accelerometer and gyroscope, which can measure linear acceleration and angular velocity of heads. The mouthguard can rigidly fit to the upper dentition, therefore sensors can accurately measure the head kinematics \cite{27}. All head impacts recorded by the instrumented mouthguard were confirmed by videos, and in total 118 head impacts were included in the dataset \cite{19}. 

To obtain the impact locations of these head impacts, the screenshot of the head impact was saved and reviewed by five reviewers who have experience in analyzing American Football head impacts. According to the screenshot, five reviewers determined if the impact location is "facemask", "Top of helmet", "Back of helmet", "Left of helmet" or "Right of helmet" independently. If more than three reviewers observed a head impact at the same location, we considered this impact location of this head impact as clearly identified. Otherwise, the head impact was excluded from the dataset. Then, the rest of the head impacts (n=79) were used as a reference dataset to test the ability of the impact retrieval model as well as previous models \cite{22,23,24} to determine the location of head impacts. The impact retrieval model as well as previous models calculate the spherical coordinate of the impact location $\eta$ and $\theta$, which were different from the description of the impact location in video reviews. To transform $\eta$ and $\theta$ to the impact location description, we measured the FE model of the Riddell helmet, and determined the head impact with $\eta<$ -34° is at the "Top of helmet". Otherwise, if -45°$<\theta<$ 45°, the impact is at "Facemask"; 45°$<\theta<$ 135°, it’s at "Right of helmet"; -135°$<\theta<$ -45°, it’s at "Left of helmet"; 135° $<\theta<$ 180° or -180°$<\theta<$ -135°, it’s at "Back of helmet". Then, the ratios of correct predictions were compared among different methods to show their ability to determine the impact locations.

\subsection{Impact Retrieval Deep Learning Model}
Head kinematics are results of the impact force, so it is possible to predict the impact information based on the head kinematics. We developed the force retrieval model to retrieve the head impact information including the speed, orientation parameters ($\alpha$, $\beta$, Y, and Z), as well as the impact force curve according to the head kinematics. Recurrent Neural Networks (RNNs), and particularly Long Short-Term Memory (LSTM) networks, have gained prominence in analyzing time-series data and signal processing. In this study, we harnessed LSTM-based RNNs to infer impact speed, orientation and force profiles from head kinematic data, using a training dataset derived from Riddell helmet head impact simulations.

The input of the force retrieval model is head kinematics, which consists of signals spanning 145 milliseconds, capturing the duration of the impact. We enhanced model accuracy by incorporating a rich set of features: tri-axial angular velocity and its magnitude (4 channels), tri-axial angular acceleration and its magnitude (4 channels), and tri-axial linear acceleration and its magnitude (4 channels). These were captured across four reference frames: global, global spherical, local, and local spherical, culminating in 48 signal channels (12 channels of kinematics under four reference frames), each with a 145 ms timeframe.

The output of the force retrieval model includes head impact parameters: $\alpha$, $\beta$, Y, Z, and the impact speed, as well as the force profiles exerted on the helmet and head/face (in kN). The model outputs a single value for each of the first five metrics and a 145-element vector for the force profiles.

To strike a balance between model complexity, the risk of overfitting, and computational efficiency, our LSTM architecture comprised two one-dimensional LSTM layers, each succeeded by a dropout layer, and a fully connected dense layer preceding the output layer. 

The LSTM's hyperparameters—including the number of units per LSTM cell, learning rate, number of epochs, dropout rate, and L2 regularization on the LSTM kernel—were meticulously optimized on a validation dataset to minimize the Mean Absolute Error (MAE). For force profile predictions, both LSTM layers were configured to return sequences, while for the impact information (speed and orientation parameters) prediction, only the first LSTM layer returned sequences. Therefore, assuming the batch size of training data was N while the hidden units were D, for the impact information prediction model, the shapes of data flow were: Input: (N, 145, 48), LSTM layer 1 output: (N, 145, D), Dropout layer 1 output: (N, 145, D), LSTM layer 2 output: (N, D), Dropout layer 2 output: (N, D), Final output: (N, 1). For the force profile prediction, the shapes of the data flow were: Input: (N, 145, 48), LSTM layer 1 output: (N, 145, D), Dropout layer 1 output: (N, 145, D), LSTM layer 2 output: (N, 145, D), Dropout layer 2 output: (N, 145, D), Final output: (N, 145, 1). This approach was designed to ensure the LSTM models were adequately expressive yet computationally manageable, allowing for accurate and efficient predictions of both impact characteristics and force profiles.

To rigorously assess the performance of our model and the robustness of our test results, we distributed the comprehensive Riddell helmet head impacts dataset into three distinct subsets: training, validation, and testing, with an allocation of 80\%, 10\%, and 10\%, respectively. This partitioning was replicated across 20 parallel experiments, each initialized with a unique random seed to test the robustness of the models. For each subset within these partitions, we employed three key metrics to quantify the accuracy and predictive power of our model: Mean Absolute Error (MAE), Root Mean Squared Error (RMSE), and the Coefficient of Determination ($R^2$). These metrics were calculated for both validation and test datasets to ensure thorough evaluation. Specific to the task of force profile prediction, we undertook a more granular analysis. We calculated pointwise MAE and RMSE to assess overall fidelity at each timestep, and peak MAE, peak RMSE, and peak $R^2$ to specifically scrutinize the model's accuracy at critical points of impact.

The outcomes of these 20 parallel experiments were synthesized to provide a comprehensive overview of the model's performance. In the results section, we have presented these findings in the form of summary statistics—namely the mean and standard deviation for each metric—to offer a clear and statistical insight into the model's predictive reliability.

\section{Results}
\subsection{Predicting Head Impact Parameters in in-silico Head Impacts Dataset}
\begin{figure}
    \centering
    \includegraphics[width=\linewidth]{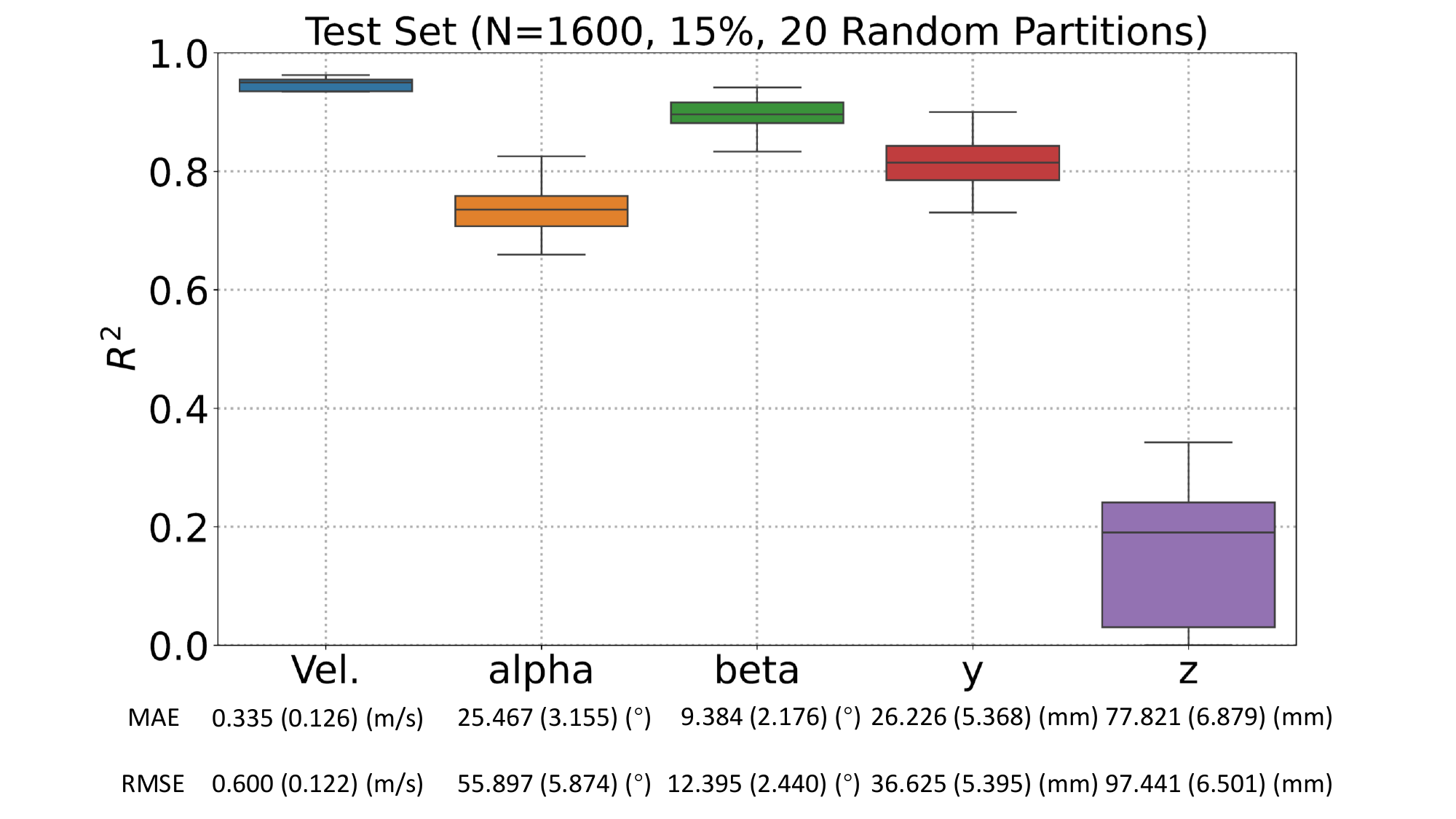}
    \caption{The $R^2$ of predicting impact speed and orientation based on head kinematics based on the LSTM model on the Riddell Helmet finite-element-simulation datasets. The mean and standard deviation of mean absolute error (MAE) and root mean squared error (RMSE) on the test set over 20 experiments with random training/validation/test dataset partitions were reported.}
    \label{fig2}
\end{figure}

To evaluate the potential of estimating impact speed, orientation and force using head kinematic signals captured, we developed LSTM models utilizing the Riddell Helmet finite-element-simulation datasets (n=16,000). After training on 80\% of the data, the model demonstrated high accuracy when assessed against the 10\% hold-out test data (n=1,600). Details of the tuned hyperparameters can be found in Table S1. Fig. \ref{fig2} shows the performance of the force retrieval model in terms of head impact parameter prediction and promising $R^2$ can be found in predicting the impact parameters except for Z. The LSTM models achieved coefficients of determination exceeding 70\% for all tasks except for the prediction of Z. The mean absolute error (MAE) values, computed from 20 parallel experiments for estimating impact speed, $\alpha$, $\beta$, and Y, were 0.335 m/s, 25.467 degrees, 9.384 degrees, and 26.226 mm, respectively.

\subsection{Predicting Head Impact Force in in-silico Head Impacts Dataset}
\begin{figure*}

    \centering
    \includegraphics[width=0.9\linewidth]{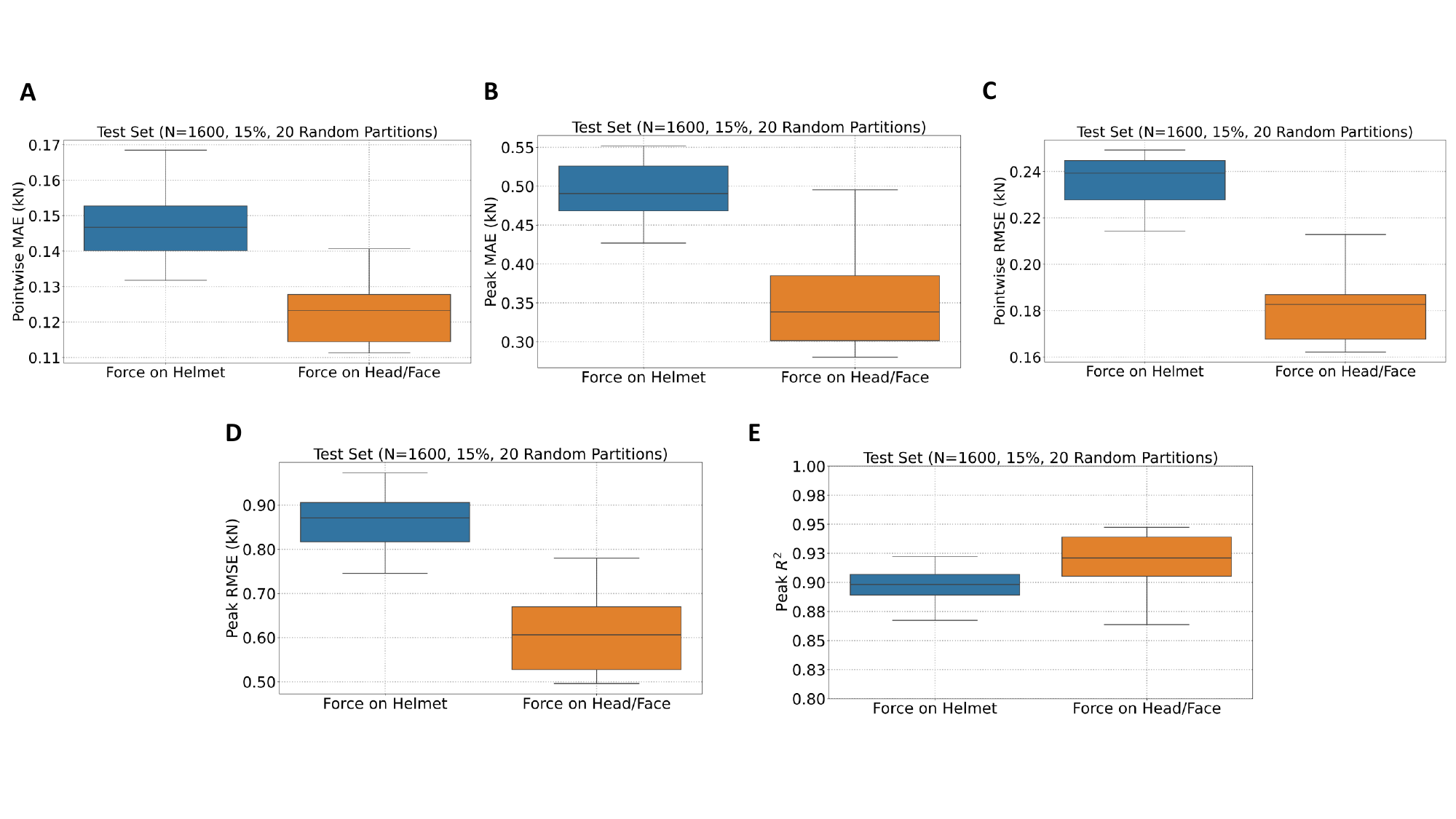}
    \caption{The performance of predicting force profiles based on head kinematics based on the LSTM model on the Riddell Helmet finite-element-simulation datasets. The distribution of the performance metrics on the test set over 20 experiments with random training/validation/test dataset partitions were reported: pointwise mean absolute error (MAE) (A), peak MAE (B), pointwise root mean squared error (RMSE) (C), peak RMSE (D), peak $R^2$ (E).}
    \label{fig3}
\end{figure*}

\begin{figure*}
    \centering
    \includegraphics[width=0.85\linewidth]{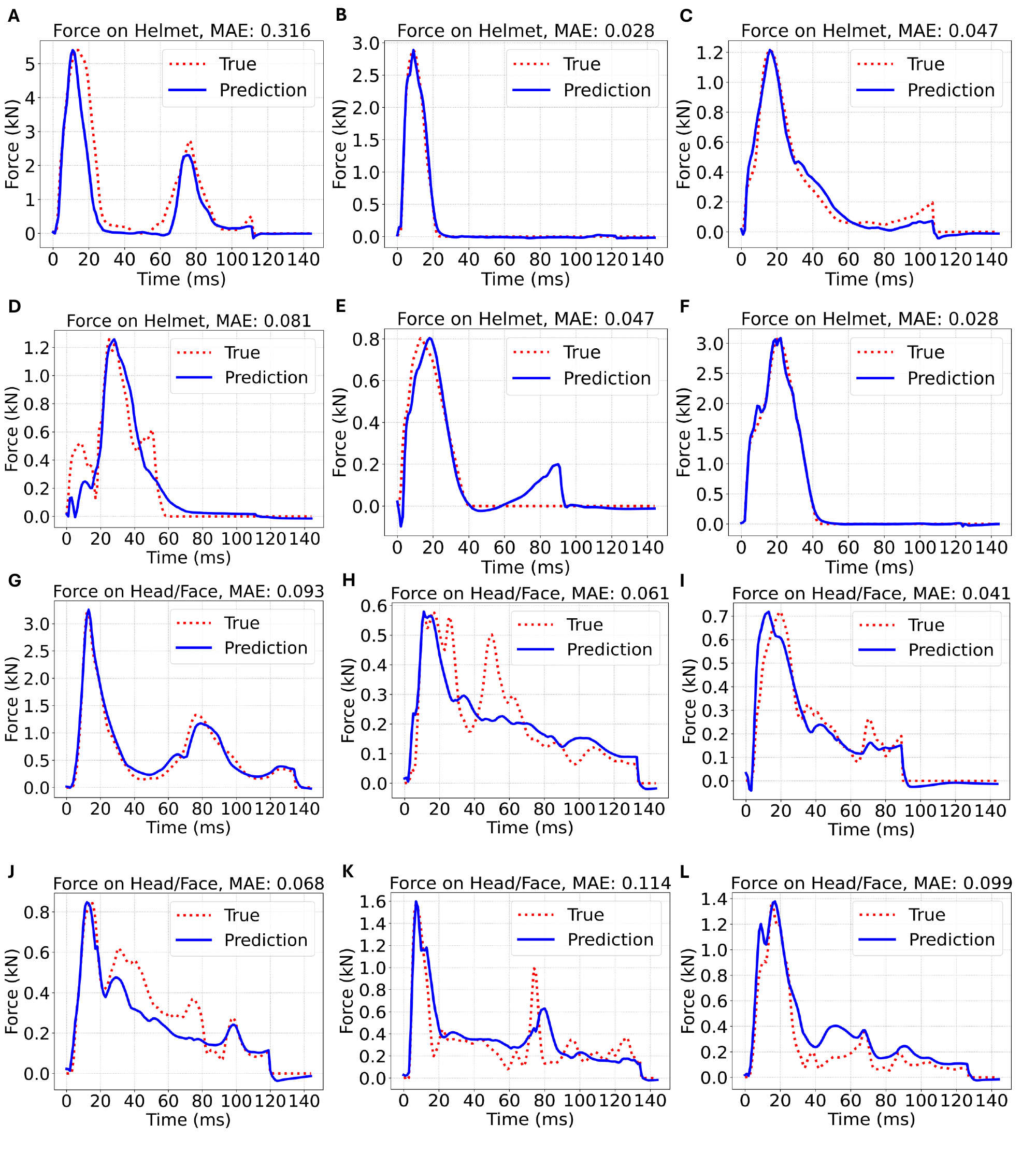}
    \caption{The examples of force profile prediction are based on the LSTM model on the Riddell Helmet test set. The reference results given by finite-element simulation and LSTM prediction were shown. The examples for force on the helmet (A-F), for force on the head/face (G-L) are selected based on one randomly selected partition of training/validation/test partition among 20 repeated experiments. The pointwise MAE values were labeled in the subplot titles.}
    \label{fig4}
\end{figure*}

Besides the prediction of impact parameters, we also developed LSTM models to predict the impact force. Fig. \ref{fig3} shows the performance of the force retrieval model in terms of impact force profile prediction. We assessed the pointwise MAE, pointwise RMSE, peak MAE, peak RMSE, and peak $R^2$ on the hold-out test set (n=1,600) in 20 parallel experiments with random test set partitions. Both the pointwise MAE and peak MAE measured less than 0.15kN and 0.5kN, respectively, while the peak $R^2$ values exceeded 85\%. To offer a visual representation of the prediction accuracy, we randomly selected one test set among 20 parallel experiments and showcased 12 examples in Fig. \ref{fig4}. These visual results and statistics corroborate that the LSTM models typically provide precise predictions of force profiles, capturing both peak values and overall traces accurately.

To sum up, the LSTM models demonstrated promising accuracy in estimating impact speed, orientation and force profiles using the Riddell helmet datasets (n=16,000), particularly with an extensive training set (n=12,800). 

\subsection{Predicting Head Impact Location in on-field American Football Dataset}
\begin{figure*}
    \centering
    \includegraphics[width=0.95\linewidth]{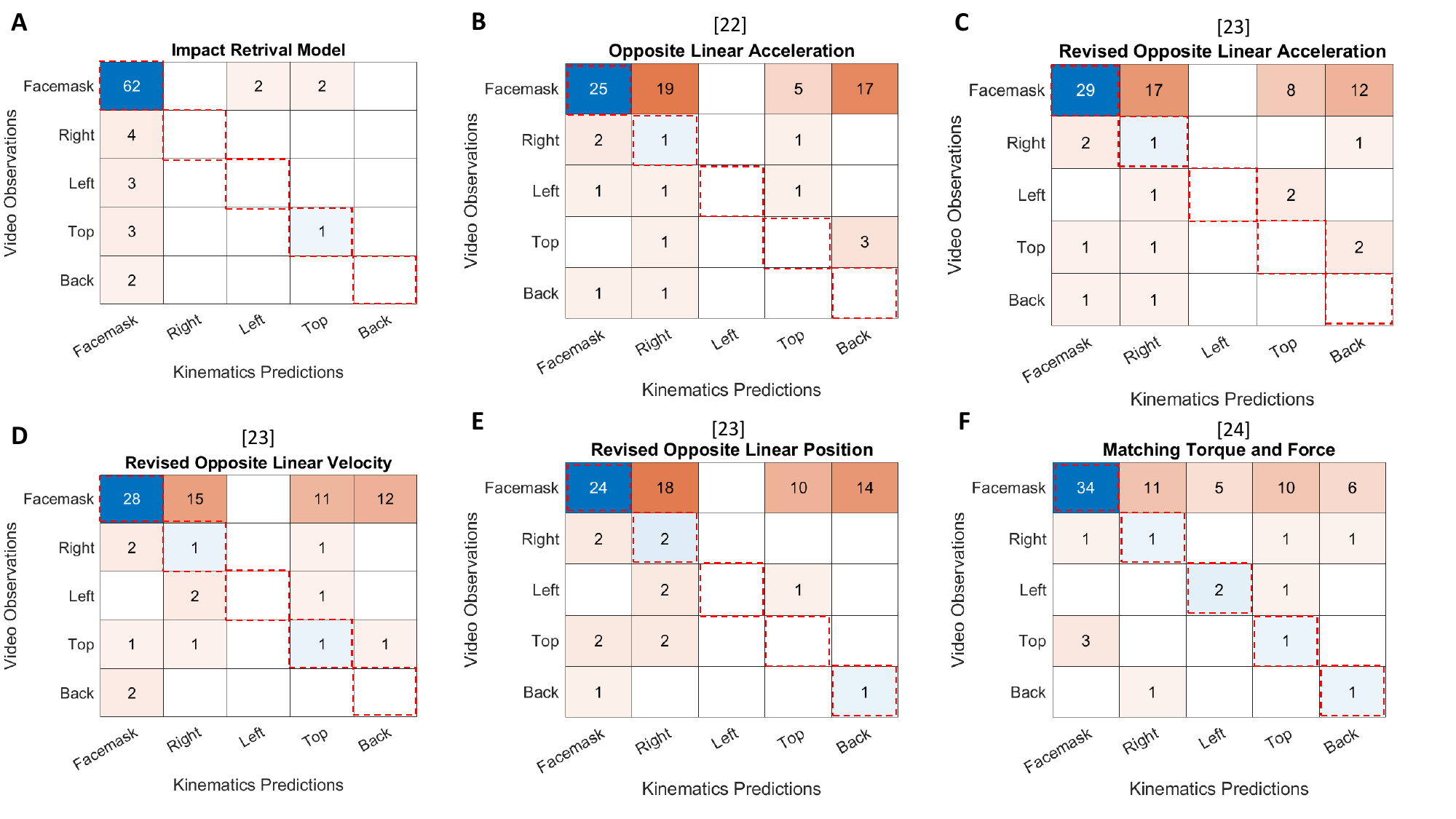}
    \caption{The confusion matrices in impact direction identification on the video-verified on-field American football dataset from different models. The reference for 79 impacts was given by two independent video reviewers. Orange: wrong predictions, blue: correct predictions, emphasized with dashed red boxes. The confusion matrices of the impact retrieval model developed in this study (A) show a clear improvement in identification accuracy when compared against the existing methods (B-F).}
    \label{fig5}
\end{figure*}

To assess the real-world applicability of impact parameter prediction models, specifically their accuracy, we employed impact retrieval models (LSTM) to predict the head impact information and then to identify the impact locations of on-field American football head impacts in which the impact locations can be clearly observed. The impact locations identified by the models were compared with the ones observed from the video, which were used as references to evaluate the ability of models. Furthermore, the methods developed in previous studies \cite{22,23,24} were also compared with the references. The outcomes, illustrated in Fig. \ref{fig5} via confusion matrices, highlight the superior identifying accuracy of our LSTM-based impact retrieval model. It successfully identified 63 out of 79 impacts, achieving an accuracy rate of 79.7\%, the highest among the evaluated methods. This performance surpasses that of the revised opposite linear velocity and acceleration methods (successfully identified 29 out of 79 impacts, achieving an accuracy of 36.7\%) proposed by Kuo et al. \cite{23}, and that of the matching torque and force method  (successfully identified 39 out of 79 impacts, achieving an accuracy of 49.4\%) proposed by Bartsch et al. \cite{24} These results show that our impact retrieval models trained on the finite element simulated impacts can be more accurate in retrieving the impact direction when compared with the existing methods.

\section{Discussion}
The goal of this study is to develop a method to accurately reconstruct the head impact information including the impact speed, location, direction, and impact force, which are essential to the development of helmet and head impact data analysis. Currently, massive head impact kinematics data in sports has been collected by wearable inertial sensors, and the head impact information was estimated by methods based on mechanical simplifications and assumptions. However, estimating is difficult because the head impact process is complicated. First, when the impact mass contacts the helmet, the impact is mitigated by the energy-absorbing system, and the impact force is transmitted to the head through the interaction between the helmet and the head. This interaction includes compressional force that is normalized to the head surface and the frictional force that is tangential to the head surface. Therefore, the direction of the impact force may change when it is transmitted to the head. Second, the head rotates because of the impact force and the constraint of the neck, and the instantaneous center of head rotation varies in sagittal extension, while it was fixed in the coronal flexion. The previous methods to estimate the head impact information simplified the impact process significantly \cite{23}, which inhibited them from predicting the impact speed and force. The impact location and direction can be predicted, but the accuracy is not satisfying. 

To overcome the difficulty brought by the complexity of the head impact process, we developed a head impact retrieval model to predict the head impact information. Recurrent neural networks are particularly useful in modeling time series and signal-based data. LSTM, as a type of recurrent neural networks, was adopted considering its ability to learn the long-term temporal dependencies relationship: the head kinematics are dependent on the whole impact process instead of the instantaneous impact force. Additionally, different from the conventional feed-forward neural networks, the LSTM doesn’t require feature engineering to extract relevant information from the time traces based on expert knowledge \cite{28,29}, and it learns the feature representation in a completely data-driven manner. To train the model , we generated a dataset using FE model \cite{26} because it can provide a large amount of head impacts and the numerical simulation can provide all the head impact information. The prediction of impact speed is the most accurate, potentially because the impact speed is closely related to the magnitude of the head acceleration. $R^2$ of the predictions for $\beta$ and Y are higher than 0.8,  while $R^2$ of the prediction for $\alpha$ is 0.737. These show the ability of this impact retrieval model to reconstruct the head impact. However, the impact retrieval model could not accurately predict Z ($R^2$ of the prediction for Z is only 0.159). This may be owing to the relative position between the impact location and the instantaneous head rotation axis. For different Y, the impact location varies from left to right of the rotation center, and the head kinematics are in the opposite direction. However, for different Z, the impact location is always higher than the rotation center, so the head kinematics are closer, which makes predicting Z from the kinematics more difficult. Besides the head impact parameters, the retrieval model can accurately predict the impact force (Fig. \ref{fig3}). The MAE of both force on the helmet and force on the head/face are smaller than 0.2 kN, and the $R^2$ for the peak force is close to 0.9. 

The on-field American football head impact dataset was used to test if the impact retrieval model can be translated to the real head impact. Based on the video, each head impact was labeled by one of five different locations, which were used as the reference. Then, the impact location predicted by the retrieval model as well as previous methods were compared with the reference. 79.7\% of head impact locations were correctly identified, which is significantly higher than the previous methods. The second best is the matching force and torque method proposed by Bartsch et al in 2014 with an accuracy of 49.4\%. We also found the revised opposite linear acceleration gave the most accurate prediction than linear velocity and position, which agrees with the comparison reported by Kuo et al. \cite{23}. The ratio of correct identification is 37.6\% for the revised linear acceleration model, which is close to the ratio reported in the paper 37.3\%. It should be noted that most of the head impact was facemask head impact, and the accuracies of other methods were limited because they mispredicted some of the facemask head impacts as side impact or back impact. Since the impact direction was assumed as the direction of the linear acceleration of the head, a potential explanation for why the previous method cannot accurately predict the facemask head impact is that the interaction between the helmet and the head and the energy absorbing system of the helmet change the direction of the impact force.

This study has several limitations: First, the difference in the helmet has not been adopted. Both in-silicon and on-field datasets were collected with American Football helmets. The energy absorption varies across different helmets, which affects the relationship between the head kinematics and the head impact information. This difference may produce considerable errors when the current retrieval model is used with other helmets. This limitation can be solved by adopting a transfer learning technique, which requires a smaller dataset to tune the current retrieval model; Second, most of the head impacts in the on-field data head impact dataset are facemask impacts. Although this agrees with the distribution of the impact location in American Football games, the numbers of other head impacts were small, therefore the ability of the retrieval model to detect other head impact locations is not well tested. This can be solved with more head impact datasets in the future.

\section{Conclusion}
This study introduces a novel deep learning approach to accurately predict head impact parameters, including location, speed, orientation, and force, based on helmeted head kinematics. By utilizing a Long Short-Term Memory (LSTM) network trained on a comprehensive dataset of 16,000 simulated helmeted head impacts, we have demonstrated the model's ability to achieve $R^2$ values exceeding 70\% across all predictive tasks. Furthermore, validation against an on-field dataset of 79 recorded head impacts showcased the model's superior performance, with a remarkable accuracy of 79.7\% in identifying impact locations, significantly outperforming traditional methods. The findings underscore the significant potential of our model in enhancing helmet design and safety in sports by providing more precise and reliable impact data.

\section{Code and Data Availability}
The code to develop this model and the simulation datasets are posted at: https://github.com/xzhan96-stf/impact\_retriever.git.

\section{Declaration of Interest}
The authors declare no conflicts of interest.

\ifCLASSOPTIONcaptionsoff
  \newpage
\fi

\bibliographystyle{IEEEtran} 
\bibliography{cited}

\end{document}